\documentstyle[twoside,amssymb,fleqn,espmodified]{article}

\input epsf

\newcommand{\AmS}{{\protect\the\textfont2
  A\kern-.1667em\lower.5ex\hbox{M}\kern-.125emS}}

\def\Fbp{\hbox{${\cal F}_{n-1}^{\rm (bpm)}$}}
\def\Rb{\hbox{${\Bbb R}^{\rm (boost)}$}}

\title{Solutions of arbitrary topology in 1+1 gravity}

\author{T. Kl\"osch\address{Institut f\"ur Theoretische Physik,
	Technische Universit\"at Wien, \\ 
	Wiedner Hauptstr. 8--10, A-1040 Vienna, Austria}%
        \thanks{supported by the Austrian Fonds zur F\"orderung der
	wissenschaftlichen Forschung (FWF), project P10221-PHY.}}
       
\begin{document}
\begin{abstract}
We present a classification of all global solutions for generalized 2D
dilaton gravity models (with Lorentzian signature). While for some of
the popular choices of potential-like terms in the Lagrangian,
describing, e.g., string inspired dilaton gravity or spherically
reduced gravity, the possible topologies of the resulting spacetimes
are restricted severely, we find that for generic choices of these
`potentials' there exist maximally extended solutions to the field
equations on {\em all\/} non-compact two-surfaces.

\medskip

Talk given at the Second Conference on Constrained Dynamics and Quantum
Gravity, Santa Margherita Ligure, September 17--21, 1996. To appear in the
proceedings.

\end{abstract}

\maketitle

\section{INTRODUCTION}

This talk is a short summary of \cite{III}. The restriction to space- and
time-orientable solutions without degenerate Killing horizons allows for a
condensed presentation.

\subsection{Models}

The models to be treated comprise all 2D dilaton gravity
theories, with Yang-Mills fields of an arbitrary gauge group,
\begin{eqnarray}
  \lefteqn{ L[g,\Phi,A] = \int_{\cal M} d^2 x \sqrt{|\det g|}
                            \big[U(\Phi) R + V(\Phi) + } \nonumber \\
  & & W(\Phi) \partial_\mu \Phi \partial^\mu \Phi +
                 K(\Phi) \mbox{tr}( F_{\mu \nu} F^{\mu \nu} ) \big] \, ,
  \label{model}
\end{eqnarray}
thus including e.g.\ spherically reduced gravity, the Jackiw-Teitelboim
model, or $R^2$-gravity, but also generalizations with non-trivial torsion.
In all of the above cases the solutions could locally
be brought into Eddington-Finkelstein form \cite{I},
\begin{equation}
  g=2drdv+h(r)dv^2 \, ,
 \label{011h}
\end{equation}
the dilaton $\Phi$ and all other physically relevant fields depending on
$r$ only. This form of the local solutions is equivalent to the
existence of a local Killing symmetry (generator $\frac\partial{\partial v}$).
It will be the only necessary ingredient for our analysis.

For a fixed Lagrangian (\ref{model}) a one-parameter family of functions
$h=h_M(r)$ in (\ref{011h}) is obtained, where $M$ can often be given a
physical interpretation as black-hole mass; if YM-fields are present,
furthermore, then the YM-charge $q$ of the quadratic Casimir enters
$h$ as a second parameter.

The issue of this talk can thus be stated as follows: given a local solution
(\ref{011h}) in terms of $h(r)$, what global solutions arise?

\subsection{Anticipation of the results}

The classification will be found to depend solely on the number and order
of the zeros of the function $h(r)$ in (\ref{011h}). Here we restrict
ourselves to simple zeros (zeros of higher order occur only for specific
values of the parameter $M$).
Besides the uniquely defined universal covering one obtains the following
global space- and time-orientable solutions ($n=\mbox{\# simple zeros}$):
\begin{list}{}{\leftmargin1.2cm\labelwidth1cm\itemindent.0cm\labelsep.2cm}
\item[$n=0$:] Cylinders labelled by their circumference (real number).
\item[$n\ge1$:] The above cylinders are still available, but
  now incomplete in a pathological manner (Taub-NUT space).
\item[$n=2$:] Complete cylinders labelled by a discrete parameter
  (patch number) and a further real parameter.
\item[$n\ge3$:] Non-compact surfaces of arbitrary genus with
  an arbitrary number ($\ge1$) of holes. The number of continuous parameters
  equals the rank of the fundamental group $\pi_1(\mbox{solution})$, and
  there are also further discrete parameters.
\end{list}
The solution space for a fixed model (\ref{model}) and fixed topology is
labelled by the above parameters {\em plus\/} the mass parameter $M$. For the
case of additional YM-fields its dimension generalizes to
$(\mbox{rank }\pi_1(\mbox{solution})+1)(\mbox{rank (gauge group)}+1)$.

\section{TOOLS}

\subsection{Method}
\label{method}

Our treatment is based on two standard mathematical theorems:
\begin{quote}
{\bf Theorem 1:} All smooth {\em global\/} solutions are obtained by factoring
the universal covering $\cal M$ by a freely and properly discontinuously
acting subgroup $\cal H$ of the symmetry group $\cal G$.
\end{quote}
\begin{quote}
{\bf Theorem 2:} Two factor spaces ${\cal M}/{\cal H}$ and
${\cal M}/{\cal H}'$ are isomorphic, iff the subgroups are conjugate,
i.e.\ ${\cal H}'=g{\cal H}g^{-1}$ for some $g\in{\cal G}$.
\end{quote}
By {\em global\/} we mean that the solutions are maximally extended in the
sense that the extremals are either complete at the boundary, or some
physical field (curvature $R$, dilaton field $\Phi$) blows up there, rendering
a further extension impossible. There are other examples of inextendible
solutions, where conical singularities or loss of the Hausdorff property
impede an extension, but they shall not be considered here 
(cf.\ \cite{III}).

\subsection{Universal Coverings}

While for $n=0$ the EF-coordinates (\ref{011h}) cover the whole spacetime, for
$n\ge1$ they are incomplete and have to be extended. This can be done by a
simple gluing procedure as described in \cite{II}, leading to the Penrose
diagrams of Fig.\ 1. For $n\ge2$, however, these {\em basic patches\/}
are still incomplete and have to be extended by appending similar patches
along the shaded faces. The number of pairs of these faces is $n-1$.

The thin interior lines of Fig.\ 1 are Killing trajectories ($\Phi=const$),
the dashed lines (including the shaded ones) mark Killing horizons
(zeros of $h$).
Note that these are only special examples: sectors within a basic patch may be
rearranged, for different functions $h$ the triangular sectors might have to be
replaced by square-shaped ones or vice-versa, or the whole patch might be
rotated by $90^\circ$ (like in Fig.\ 5), without affecting our
analysis, certainly.
\begin{figure}[h,t]
\epsfxsize 7.5cm \epsfbox{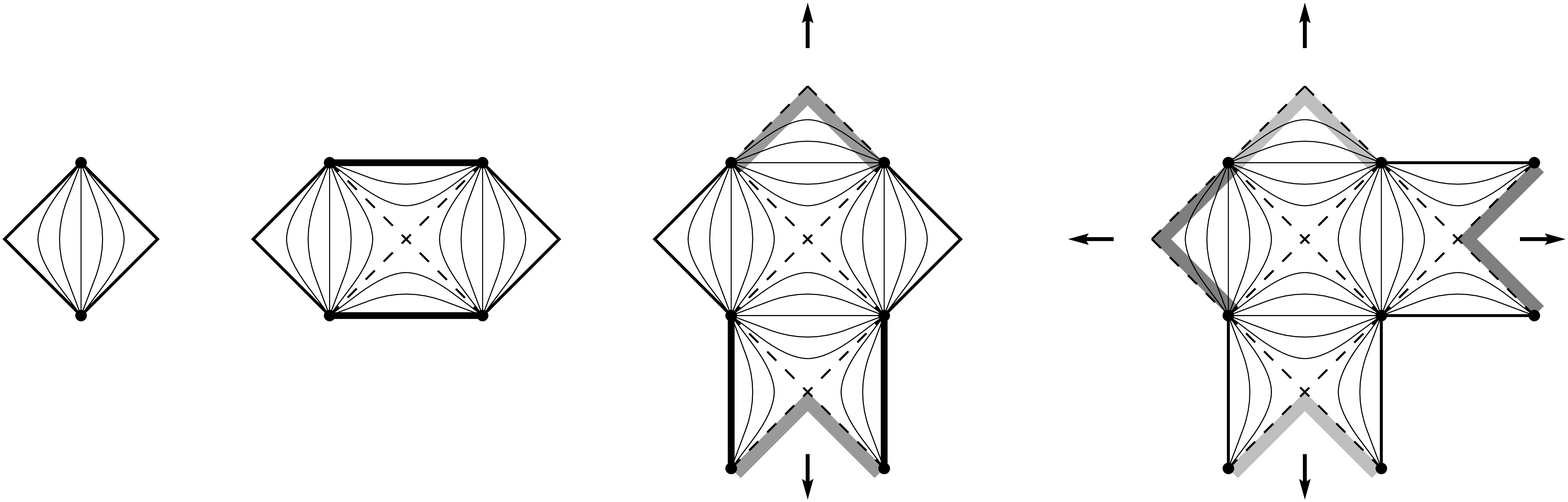}
\renewcommand{\baselinestretch}{.9}
\small \normalsize
\vspace{-.5cm} 
\caption{\small Basic patches for $n=0,1,2,3$.}
\end{figure}

\subsection{Symmetry group}

The symmetries have to preserve the dilaton field $\Phi$, metric, and
orientation, of course. The resulting group turns out to be a direct
product of a free combinatorial group \Fbp, and the group $\Rb\cong\Bbb R$,
\begin{equation}
  {\cal G}=\Fbp\times\Rb\,.
 \label{Sym}
\end{equation}
We will denote its elements by $(g,\omega)$.
\Fbp\ corresponds to permutations of the basic patches as a whole
(`basic-patch-moves', or `bp-moves') and is generated by moves across an
incomplete (shaded) face of the basic patch to the adjacent one
(cf.\ Fig.\ 2). Thus its rank is $n-1$.
\begin{figure}[h,b,t]
\epsfxsize 5cm \epsfbox{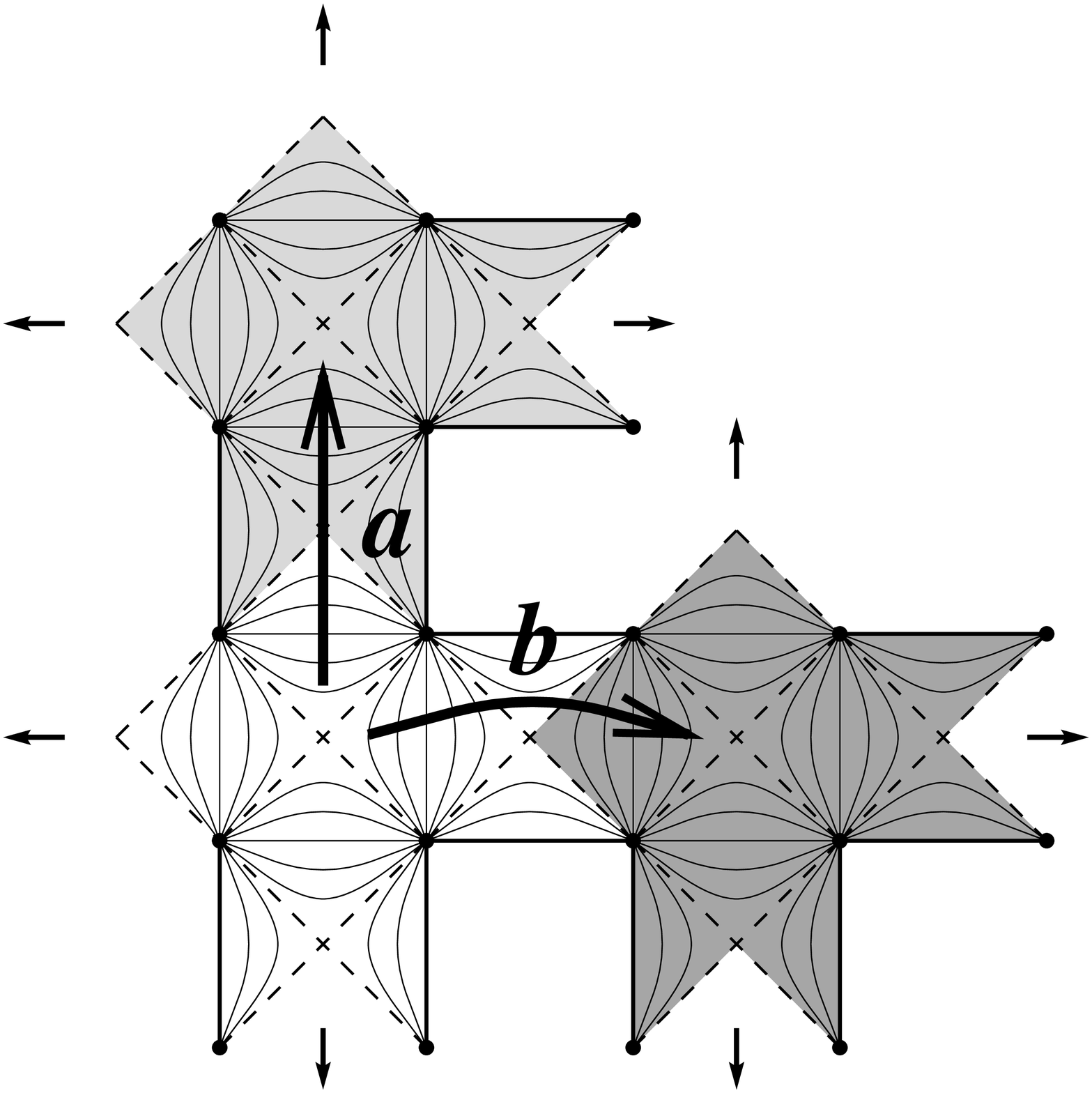}
\renewcommand{\baselinestretch}{.9}
\small \normalsize
\vspace{-.5cm}
\caption{\small The two generators of \Fbp\ for $n=3$.}
\end{figure}
The second factor, \Rb, contains the Killing transformations (boosts).
In EF-coordinates (\ref{011h}) such a boost is a shift of length $\omega$ in
$v$-direction. Its action on a basic patch is 
sketched in Fig.\ 3.
\begin{figure}[t]
\epsfxsize 3cm \epsfbox{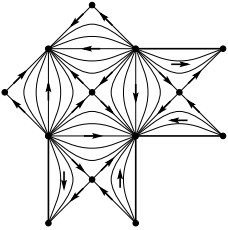}
\renewcommand{\baselinestretch}{.9}
\small \normalsize
\vspace{-.5cm}
\caption{\small Killing-field generating a boost.}
\end{figure}

\section{FACTOR SPACES}

According to \ref{method} all solutions are obtained as factor spaces of the
universal covering by an adequate group action. We have thus to
find the conjugacy classes of subgroups ${\cal H}\le{\cal G}$.
For illustrative purposes we will also use the alternative description by
fundamental regions (i.e.\ subsets of the universal covering, which cover
the factor space exactly once). The group action is then encoded in the way
the incomplete boundary faces of the fundamental region have to be glued.

\subsection{Factor spaces from boosts}
\label{pure}

The only discrete subgroups of boosts are the infinite cyclic groups
generated by one boost,
${\cal H}_\omega := \left\{(id,k\omega),\, k\in{\Bbb Z} \right\},\: \omega>0$.
The resulting factor space can be described easily for $n=0$, where the
EF-coordinates (\ref{011h}) cover the entire spacetime:
as a fundamental region one can choose a strip of width $\omega$ in the
$v$-coordinate; the factor space is then clearly a cylinder.
Also, since \Rb\ is abelian and a direct factor in $\cal G$, 
the group ${\cal H}_\omega$ is invariant under conjugation
and the parameter $\omega$ cannot be changed. Thus the
cylinders are labelled by one positive real parameter $\omega$,
which has a natural geometric interpretation in terms of the metric-induced
circumference of a fixed $\Phi=const$-line.

However, for $n\ne0$ the action of ${\cal H}_\omega$ is not properly
discontinuous at the Killing horizons and the factor space consequently
not Hausdorff (Taub-NUT spaces). [When restricting to one EF-patch
(\ref{011h}) only, then the above construction yields regular cylinders
indeed, but they are no longer complete.]
If such pathological solutions are to be excluded, then the subgroups
have to be restricted considerably: no pure boosts are allowed, but also 
there must not occur the same bp-move twice with different boost-parameters,
$(g,\omega_1)$ and $(g,\omega_2)$, as they could be combined to a pure
boost, $(id,\omega_2-\omega_1)$.

Thus one is lead to the following strategy: `forget' in the first place the
boost-component and consider $\cal H$ as subgroup of \Fbp\ only. Since
subgroups of free groups are again free, this analysis can be carried out
explicitly. Only afterwards re-provide the generating bp-moves with their
boost-parameters.

\subsection{Factor spaces from bp-moves}

For $n=0,1$, \Fbp\ is trivial and thus there are no factor spaces besides
the above boost-cylinders. For $n=2$, on the other hand, it has one generator
(e.g.\ in Fig.\ 5 a shift one patch sidewards). Consequently there occur
cylinders labelled by their (integer) patch number. For an interpretation of
the (real) boost-parameter cf.\ Sec.\ \ref{intpr}.

More complicated topologies may be obtained for $n\ge3$ (at least two
generators). Note that for the fundamental group of the factor space we have
$\pi_1({\cal M}/{\cal H})\cong{\cal H}$. Thus, if the number of holes is known
(by counting the connected boundary components, cf.\ Fig.\ 4), then
\begin{equation}
  \mbox{genus}\,=\,\frac{\mbox{rank}\,{\cal H}-(\mbox{\# holes})+1}2 \,.
 \label{genus1}
\end{equation}
\begin{figure}[h,t]
\epsfxsize 7.5cm \epsfbox{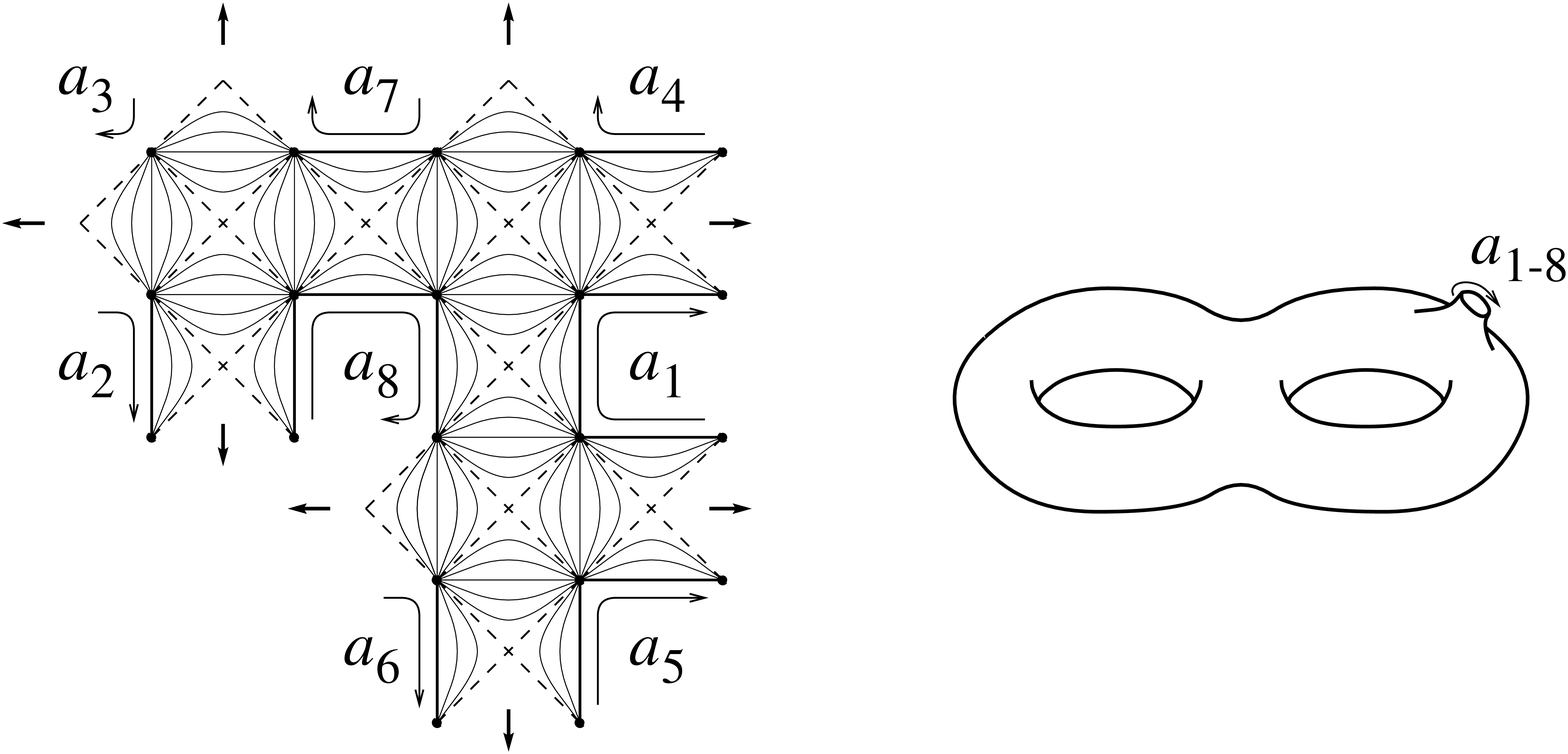}
\renewcommand{\baselinestretch}{.9}
\small \normalsize
\vspace{-.5cm}
\caption{\small Genus-2-surface with hole for $n=3$. A fundamental region
  (3 patches) is depicted on the left, where opposite faces are to be glued
  together. The eight boundary segments $a_{1\!-8}$ constitute a single hole.}
\end{figure}

By means of explicit examples it can be shown that indeed solutions of
arbitrary genus with an arbitrary nonzero number of holes (both may be
infinite) occur. Furthermore, for a finite number of basic patches within
the fundamental region, one can dispense with determining
$\mbox{rank}\,{\cal H}$ explicitly and use
\begin{eqnarray}
  \lefteqn{\mbox{genus}\,=\,\frac{(\mbox{\# patches})\cdot(n-2)-
                     (\mbox{\# holes})}2+1}            \nonumber\\  
 \label{genus2}
\end{eqnarray}
instead (cf.\ Fig.\ 4).

Each free generator of $\cal H$ carries a boost-parameter. This does not
necessarily imply, however, that the space of boost-parameters is
${\Bbb R}^{\;\!{\rm rank}\:\!{\cal H}}$: there might be a non-trivial discrete
action of ${\cal NH}/{\cal H}$ on ${\Bbb R}^{\;\!{\rm rank}\:\!{\cal H}}$
($\cal NH$ = normalizer of $\cal H$ in \Fbp); the true parameter space is then
the factor space under this action.

\subsection{Geometrical interpretation of the boost-parameter}
\label{intpr}

As already mentioned, the boost-parameters $\omega_i$ of the generators
cannot be conjugated away fully and are thus meaningful parameters for the
factor solutions. In the case of pure boosts (\ref{pure}) there was
a nice interpretation as the size (circumference) of the resulting cylinder.
This cannot be transferred, however, to the case of
non-trivial bp-moves, where another construction is useful:

As an example let us choose an $n=2$ cylinder of two basic patches
circumference (Fig.\ 5). The generating bp-move (two patches to the
left) dictates that the rightmost basic patch has to be mapped onto the
corresponding leftmost one. A non-trivial boost-parameter implies
that a boost has to be applied during this mapping, thereby distorting
the straight line of the right patch into the curved one left.
[Thus, another possible fundamental region (besides that consisting of two
entire patches) is the shaded area in Fig.\ 5.]

In \cite{II} it was shown that the intersection points of the Killing horizons
are conjugate points, and there is thus a family of oscillating extremals
running through them (dotted lines in Fig.\ 5). Now, by the boost
also the tangents of these extremals are altered, so they return boosted
against the start, the enclosed angle being the desired geometric quantity.
\begin{figure}[t]
\epsfxsize 7.5cm \epsfbox{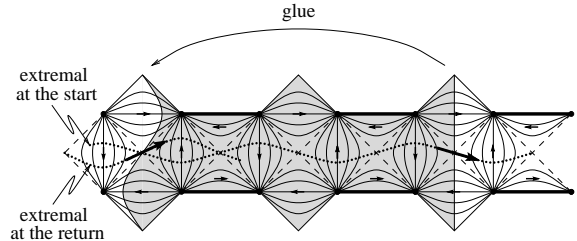}
\renewcommand{\baselinestretch}{.9}
\small \normalsize
\vspace{-.5cm}
\caption{\small Fundamental region (shaded) and interpretation of the
  boost-parameter for $n=2$. The straight right boundary of the shaded
  region has to be glued to the curved left one, and also the extremal
  (dotted line) returns tilted (i.e.\ boosted).}
\end{figure}

This construction can of course be generalized to the more complicated
solutions for $n\ge3$ (like Fig.\ 4), if necessary by considering a
polygon of mutually orthogonal extremal segments.

\section{FURTHER DEVELOPMENTS}

The above results can be generalized to non-orientable solutions as well
as to degenerate horizons (higher order zeros of $h$). This requires at the
worst two additional semi-direct factors ${\Bbb Z}_2$ in $\cal G$
(cf.\ \cite{III}). The Hamiltonian quantization of the theory will be
summarized in \cite{Strobl}. Let us mention here that the boost-parameter
$\omega$ (for topology $S^1\times\Bbb R$)
is the second coordinate besides the mass-parameter $M$ for
the reduced phase space, and that the discrete `patch numbers' for the $n\ge2$
cylinders (e.g.\ Fig.\ 5) enter the quantum mechanical wave
functions as additional discrete labels.

\end{document}